\documentclass[sigconf]{acmart}
\AtBeginDocument{%
  }

\usepackage{listings}
\usepackage{tabularx}
\usepackage{tikz}
\usetikzlibrary{shapes.geometric, arrows}
\usetikzlibrary{arrows.meta}
\usepackage{algorithmic}
\usepackage{float}
\usepackage{graphicx}
\usepackage{textcomp}
\usepackage{xcolor}
\usepackage{blindtext}
\usepackage{enumitem}
\usepackage{comment}
\usepackage{hyperref}
\usepackage{listings}
\usepackage{verbatim}
\usepackage{colortbl}
\usepackage{multirow} 
\usepackage{array}
\usepackage{makecell}
\usepackage{subcaption}
\usepackage{booktabs}
\usepackage{balance}
\usepackage[listings]{tcolorbox}
\usepackage[skip=0pt, font=small]{caption}
\usepackage{enumitem}
\setlist[itemize]{leftmargin=1em, itemsep=2pt, topsep=2pt}

\definecolor{headercolor}{RGB}{0,0,0}
\definecolor{bluekeywords}{rgb}{0,0,0}
\definecolor{greencomments}{rgb}{0,0,0}
\definecolor{redstrings}{rgb}{0,0,0}

\lstdefinestyle{mystyle}{
showspaces=false,
showtabs=false,
breaklines=true,
showstringspaces=false,
breakatwhitespace=true,
escapeinside={(*@}{@*)},
commentstyle=\color{greencomments},
keywordstyle=\color{bluekeywords}\bfseries,
stringstyle=\color{redstrings},
basicstyle=\ttfamily,
numbers=left,                    
numbersep=5pt, 
tabsize=2,
keywordsprefix={@},
numberstyle=\scriptsize\color{gray},
frame=single,
belowcaptionskip=8pt,
mathescape=true,
}

\DeclareCaptionType{mybox}[Box][List of Boxes]

\newboolean{showcomments}
\setboolean{showcomments}{true}
\ifthenelse{\boolean{showcomments}}
{\newcommand{\nb}[2]{
		\fcolorbox{black}{yellow}{\bfseries\sffamily\scriptsize#1}
		{\sf\small$\blacktriangleright$\textit{#2}$\blacktriangleleft$}
	}
	
}
{\newcommand{\nb}[2]{}
	
}

\newcommand\marco[1]{\nb{Marco}{\color{teal}#1}}

\newcommand\tiziano[1]{\nb{Domenico}{\color{green}#1}}

\setcopyright{acmlicensed}
\copyrightyear{2018}
\acmYear{2018}
\acmDOI{XXXXXXX.XXXXXXX}
\acmConference[EASE '26]{Make sure to enter the correct
  conference title from your rights confirmation email}{June 09--12,
  2026}{Glasgow, United Kingdom}
\acmISBN{978-1-4503-XXXX-X/2018/06}

\begin{document}

\title{Inferring Equivalence Classes from Legacy Undocumented Embedded Binaries for ISO 26262-Compliant Testing}

\author{Marco De Luca}
\email{marco.deluca2@unina.it}
\affiliation{%
  \institution{University of Naples Federico II}
  \city{Naples}
  \country{Italy}
}

\author{Domenico Francesco \\ De Angelis}
\email{ddeangelis@micron.com}
\email{domenicofrancesco.deangelis@unina.it}
\affiliation{%
  \institution{Micron Technology, Inc}
\institution{University of Naples Federico II}
  \city{Naples}
  \country{Italy}
}

\author{Domenico Amalfitano}
\email{domenico.amalfitano@unina.it}
\affiliation{%
  \institution{University of Naples Federico II}
  \city{Naples}
  \country{Italy}
}

\author{Pasquale Cimmino}
\email{pcimmino@micron.com}
\affiliation{%
  \institution{Micron Technology, Inc}
  \city{Naples}
  \country{Italy}
}

\author{Anna Rita Fasolino}
\email{fasolino@unina.it}
\affiliation{%
  \institution{University of Naples Federico II}
  \city{Naples}
  \country{Italy}
}


\begin{abstract}
Equivalence class partitioning is a well-established test design technique mandated by safety standards such as ISO~26262 for systematic testing of safety software. In industrial practice, however, its application to legacy undocumented embedded firmware is often hindered by incomplete or outdated functional specifications.

This paper proposes a binary-level methodology for inferring output-oriented equivalence classes directly from compiled firmware, without relying on source-level annotations or external documentation. The approach combines control-flow reconstruction and guided symbolic execution to analyze individual functions and group execution paths according to indistinguishable observable behavior, including return values and output parameters. An optional post-processing step produces human-readable representations to support comprehension and documentation.

The methodology is evaluated in an industrial automotive context through a practitioner-based study assessing correctness and interpretability. Results indicate strong alignment with expert expectations and a positive perception of readability and usefulness for supporting function understanding and test design. These findings demonstrate the feasibility and practical relevance of binary-level equivalence class inference for systematic testing of legacy undocumented safety embedded software.

\end{abstract}




\maketitle

\section{Introduction}

Modern automotive embedded systems integrate increasing functionality and performance, leading to rapidly growing software complexity. Safety standards such as \textbf{ISO~26262}~\cite{ISO26262} and \textbf{IEC~61508}~\cite{IEC61508} therefore prescribe systematic verification processes and recommend black-box test design techniques, including \emph{Equivalence Class Partitioning (ECP)}~\cite{myers2004art}, to achieve adequate input-domain coverage. In particular, ISO~26262 mandates the use of systematic test design techniques such as ECP for high-ASIL software components (Part~6, Table~8)~\cite{iso26262-6}. In industrial practice, however, applying ECP is often problematic. Automotive software frequently consists of \emph{legacy firmware} developed without complete or up-to-date functional specifications~\cite{malladi2016,DavidLo2011}. Although ISO~26262-8 requires a valid functional specification for software reuse, it allows engineers to assess suitability by analyzing design and implementation artifacts when documentation is incomplete~\cite{iso26262-8}. Consequently, reverse engineering becomes necessary to recover functional behavior and enable systematic test design. These challenges are exacerbated by typical characteristics of industrial firmware, including implicit logic, compiler-specific optimizations, undocumented dependencies, and extensive build-time variability introduced through multi-level build systems and pervasive use of preprocessor directives (e.g., \texttt{\#ifdef}). Source-level analyses may therefore produce artifacts that do not correspond to deployed firmware configurations~\cite{Abdullah2018}. As confirmed by an internal industrial survey with firmware developers and test engineers, this results in a gap between safety-standard prescriptions and the practical means available for defining and maintaining equivalence classes. Prior work has explored recovering behavioral information without formal specifications through specification mining~\cite{Ammons2002,fan2024}, symbolic execution frameworks such as KLEE, S2E, and ANGR~\cite{Cadar2008,Chipounov2011,shoshitaishvili2016state}, and constraint-based reasoning approaches~\cite{Udeshi2024,weideman2021perfume,Winkelmann2023,Amadini2019}. While effective for invariant detection, test generation, and behavioral modeling, these techniques do not explicitly treat \emph{equivalence classes (EC)} as first-class testing artifacts nor directly support output-oriented equivalence partitioning as required by equivalence-class-based test design~\cite{Huang2016}. To address this gap, we introduce an automated technique for inferring ECP directly from compiled binaries of legacy undocumented embedded C software. Operating at the binary level ensures fidelity to deployed configurations and avoids inconsistencies introduced by source-level variability. The approach analyzes executables enriched with DWARF debug information~\cite{DWARF5}, reconstructs control-flow graphs, and performs guided symbolic execution, merging explored paths through constraint-based reasoning to identify inputs that produce indistinguishable observable behavior. We adopt an \emph{output-oriented} definition of ECs~\cite{Huang2016}, where inputs are equivalent if they induce identical externally observable effects, including return values and state modifications through dereferenced pointer parameters. By grouping symbolic paths that yield the same observable outcomes, the technique recovers undocumented functional boundaries that are difficult to identify manually in legacy undocumented firmware. The main contributions of this paper are: (i) an automated methodology for recovering output-oriented ECs from compiled embedded binaries targeting undocumented or legacy safety software; and (ii) an industrial evaluation assessing their correctness, readability, and usefulness. 

The remainder of the paper is organized as follows. Section~\ref{sec:survey} discusses the industrial context; Section~\ref{sec:proposedProcess} presents the methodology; Section~\ref{sec:runningExample} provides illustrative examples; Sections~\ref{sec:evaluation} report the experimental setup and results; Section~\ref{sec:lessons} presents lessons learned; Section~\ref{sec:related} reviews related work; and Section~\ref{sec:conclusions} concludes the paper.


\section{Industrial Context and Needs Elicitation} \label{sec:survey}
The methodology presented in this paper is motivated by challenges observed in an industrial automotive context, where firmware developers and test engineers must design and maintain safety test suites under ISO~26262 constraints. In such environments, equivalence class partitioning is widely recognized as a fundamental test design technique, yet its practical application is often hindered by the lack of reliable specifications and the complexity of legacy undocumented embedded software~\cite{iso26262-8, GAROUSI201814, malladi2016, qa-systems}.
To ground the proposed methodology in concrete industrial needs, we asked to conduct a qualitative needs elicitation study based on focus groups, involving twelve practitioners from Micron Technology. Participants included firmware developers and engineers involved in software testing activities, all with experience in safety systems. The focus group was moderated by two of the authors and conducted as a structured discussion session lasting approximately three hours.

The discussion addressed three main themes: (i) the role of ECP in safety-oriented test design; (ii) the difficulties of manually defining ECs; and (iii) expectations toward automated support for EC identification and documentation. Notes from the discussion were analyzed inductively by the authors, following a thematic analysis approach to identify recurring observations and consolidate them into a set of industrial needs relevant to the design of the proposed approach.

The outcomes highlight a strong demand for automation. Participants emphasized the importance of function-level ECP for systematic coverage and ISO~26262 compliance, while reporting significant difficulties in manual application due to large input spaces, implicit state-dependent conditions, and the absence of precise expected outputs in legacy firmware. In addition, interpretability emerged as a key requirement: practitioners stressed the need for clear constraints, representative input examples, and traceability to exercised conditions to support test design, review, and maintenance.

Overall, these findings reveal a gap between safety-standard requirements and the practical means available for defining and managing EC in legacy embedded software. The identified needs directly motivate the proposed methodology, which focuses on (i) automated derivation of ECs, (ii) an output-oriented notion of equivalence aligned with observable behavior, and (iii) the generation of human-readable representations to support comprehension and reuse in industrial workflows.

\section{Proposed Methodology}
\label{sec:proposedProcess}

This section presents the two-phase methodology adopted in this work, which is based on the sequential execution of two phases described in the following.
The methodology aims to abstract the functional behavior of individual functions into ECs, defined according to Huang et al.~\cite{Huang2016}, and focuses on \emph{computational functions}, i.e., functions that do not directly interact with hardware peripherals, interrupts, or timing-dependent I/O, which in industrial practice are typically exercised through integration-level or hardware-in-the-loop testing. The two phases consist of \textit{structural analysis and function classification}, followed by \textit{symbolic execution and EC generation}.




\paragraph{Phase 1: Structural Analysis and Function Classification.}

\begin{figure}[H]
    \centering
    \includegraphics[width=0.8\linewidth]{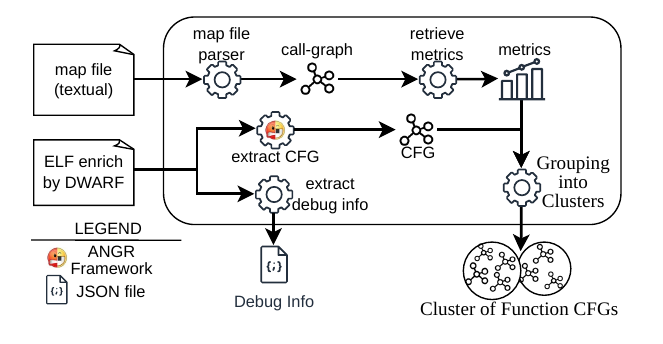}
    \caption{Phase 1 overview.}
    \label{fig:phase1}
\end{figure}

Phase~1 takes as input two build artifacts: the \emph{map file} and the \emph{executable binary} in \texttt{ELF} format~\cite{ELF}, enriched with debugging information conforming to \texttt{DWARF}~\cite{DWARF5}. The goal of this phase is to prepare and organize the analysis inputs required for Phase~2 by producing (i) clusters of function control-flow graphs (CFGs) and (ii) serialized debugging information. An overview of the phase is shown in Figure~\ref{fig:phase1}.

\color{black}

The map file is parsed to recover symbol addresses and cross-references among compilation units, enabling the reconstruction of inter-function call relationships. From the resulting call graph, two metrics are derived for each function to approximate the expected analysis effort: the \emph{call depth}, which captures the length and complexity of dependency chains, and the \emph{number of accessed global variables}, which correlates with the size of the symbolic state~\cite{Horvth2024ScalingSE}.

In parallel, the ELF binary is analyzed to extract DWARF metadata providing semantically rich information, including function signatures, parameter types, and data-structure definitions. Operating directly on the compiled binary, rather than on source code, preserves fidelity to the deployed configuration and avoids inconsistencies introduced by source-level variability. CFGs for all identified functions are then recovered from the binary using \texttt{angr}. The previously derived call-graph metrics are used to organize these CFGs into clusters, grouping functions with deeper call chains and heavier use of global state. This clustering supports more efficient downstream analysis by reducing execution time and mitigating path explosion and solver load~\cite{Horvth2024ScalingSE}. The outputs of Phase~1 are therefore: (i) a dependency-aware clustering of per-function CFGs guided by \emph{call depth} and \emph{number of accessed global variables}, and (ii) a serialized bundle of DWARF-derived debugging information for subsequent phases.

\paragraph{Phase 2: Symbolic Execution and ECP Generation.}

\begin{figure}[H] 
\centering 
\includegraphics[width=0.8\linewidth]{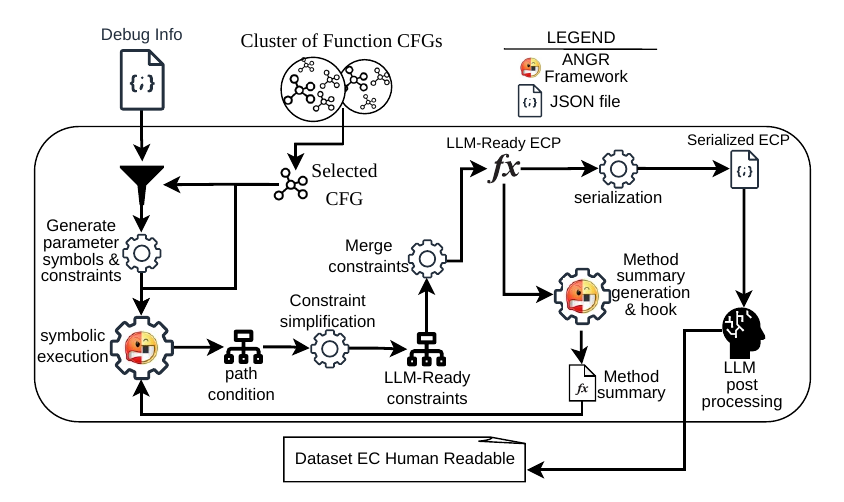} 
\caption{Phase 2 overview.} 
\label{fig:phase2} 
\end{figure}

Phase~2 derives a human-readable Equivalence Class Partitioning (ECP) from the clusters of function CFGs and the debugging information produced in Phase~1. An overview of this phase is shown in Figure~\ref{fig:phase2}. 

Clusters are processed in ascending order of call depth and, in case of ties, by increasing number of accessed global variables. This scheduling reduces symbolic execution overhead, as analyzing shallower functions first enables the construction of reusable \emph{method summaries}~\cite{yi2020summary,Zeng2024}, which are subsequently reused when exploring deeper call chains. A method summary captures the interface-level behavior of a function by mapping input conditions to observable effects, namely return values and updates to output parameters and global variables. Debug information provides the symbols required for symbolic execution and the initial assumptions over their domains. For each selected function CFG, input parameters are instantiated as symbolic bit-vectors using the \texttt{angr} framework. Type information recovered in Phase~1 is used to constrain symbolic domains, for instance by restricting Boolean parameters to $\{0,1\}$, bounding integer parameters according to their bit width, and limiting enumerated types to their declared constants. For pointer parameters, \texttt{angr}'s \emph{clinic} mechanism is used to distinguish between scalar objects and arrays, enabling bounded and accurate memory modeling. These constraints substantially reduce infeasible paths and mitigate path explosion. Symbolic execution is then performed under fixed exploration bounds. Loops are unrolled up to 1024 iterations, reflecting bounded patterns commonly observed in industrial embedded firmware, while potentially under-approximating rare corner cases~\cite{Meyer2024,Jaffar2011}. For each explored path, the analysis collects path conditions and interface-level effects, discards unsatisfiable paths, and normalizes the remaining symbolic states (e.g., last-write-wins per location and deterministic ordering of fields and elements). When an output interface yields a symbolic expression $v$, the analysis explicitly distinguishes between overflow and non-overflow cases by forking execution into $v > \texttt{MAX\_VALUE}$ and $v \le \texttt{MAX\_VALUE}$. To reduce redundancy, loop-derived paths are merged when they traverse the same loop header and exit through identical CFG nodes or edges, yielding possibly disjunctive path conditions.

Before equivalence classes are constructed, extracted path conditions undergo a pattern-based simplification step. This step rewrites recurring engine- and compilation-induced idioms into clearer and more conventional forms without altering their semantics. The simplification targets common readability issues, including fragmented byte-wise constraints, redundant Boolean structures, backend-specific operators, and other mechanically generated expressions that obscure the intent of a condition. The applied rewrite patterns are summarized in Table~\ref{tab:simplification-constraints} and are used solely as a post-processing step to facilitate human inspection and documentation~\cite{amon2020creating,Udeshi2024}. The resulting constraints are semantically equivalent to the original ones and are suitable for subsequent merging and LLM-based post-processing.

Equivalence classes are then constructed by merging symbolic states that exhibit identical interface-level behavior. For ECP purposes, paths producing the same return value and the same updates to output parameters and global variables are considered indistinguishable and are represented by a single equivalence class. Each state is characterized by its path condition and an output snapshot summarizing observable effects. Path conditions associated with identical outputs are aggregated via logical disjunction and lightly simplified to improve readability.

The merged equivalence classes are consolidated into reusable method summaries~\cite{yi2020summary,Zeng2024}. Each summary is serialized as a Python stub compatible with the \texttt{angr} framework and installed through its hooking mechanism. At call sites, the symbolic executor applies the summarized behavior directly, avoiding re-execution of the caller’s internal control flow while preserving its interface-level semantics under the same exploration bounds. This mechanism significantly reduces SMT solving and emulation costs, improving scalability on large industrial firmware. For each analyzed function, the final equivalence classes are serialized into a structured JSON representation that captures both input-domain constraints and corresponding observable outputs, enabling integration into downstream testing and verification workflows. Since symbolic constraints are expressed in low-level bit-vector logic, an optional post-processing step translates them into human-readable descriptions to support practical adoption and manual review for safety certification.

An LLM component is used only to improve the human readability of solver-validated symbolic equivalence classes, without affecting symbolic execution, class construction, or analysis decisions. It is guided by a structured prompt that frames the model as a safety engineer and asks it to translate Claripy symbolic constraints into human-readable equivalence classes. The prompt combines multiple prompt-engineering techniques from the taxonomy in \cite{prompt_eng_taxo}, enforcing a strict input/output structure, semantic preservation, and simplified, non-redundant parameter-isolated constraints with explicit input ranges. The full prompt is reported in Box~\ref{box:prompt}, while the resulting human-readable equivalence classes are stored in the \emph{Dataset EC Human Readable} file.

\newtcolorbox{fancybox}{
  colback=white,
  colframe=black,
  arc=2.5pt,
  boxrule=0.5pt
}
\begin{mybox}
    \caption{Role-constrained prompt for EC generation.}
    \fbox{\includegraphics[width=0.8\linewidth]{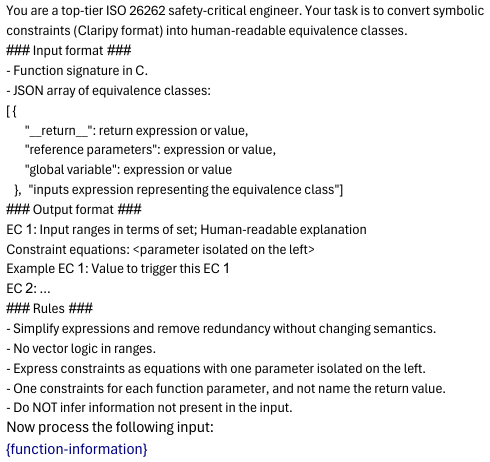}}
    \label{box:prompt}
\end{mybox}

\begin{table*}[t]
\centering
\small
\renewcommand{\arraystretch}{0.8}
\caption{Rules for constraint simplification.}
\label{tab:simplification-constraints}
\begin{tabularx}{\textwidth}{p{0.2cm} | X | p{0.8cm}}
\toprule
\textbf{ID} & \textbf{Description} 
& \textbf{Ref} \\
\midrule
1 &
When a fixed-size value is constrained by specifying constant values for individual slices, combine those slice-by-slice constraints into a single constraint on the whole value, and present the resulting constant as a signed integer (two’s complement).
& \cite{amon2020creating} \\
\hline
2 &
Simplify Boolean expressions by eliminating redundant negations/conditions, applying standard logical rewrites, and compacting overlapping comparisons on the same constant into a stronger, shorter predicate
& \cite{Udeshi2024} \\
\hline


3 &
Simplify arithmetic expressions by factoring out common multiplicative terms, reducing repetition.
& \cite{Udeshi2024} \\
\hline

4 &
Rewrite engine-specific comparison functions into their equivalent standard math comparisons to improve readability.
& \cite{Udeshi2024} \\
\hline

5 &
Detect the verbose bit-by-bit disjunction expression that encodes “at least one bit is set” and replace it with the equivalent non-zero check.
& proposed \\
\hline

6 &
Rewrite “append $k$ zeros to $x$” concatenations as a left shift $x \ll k$ to obtain a more familiar form.
& proposed \\
\hline

7 &
Rewrite comparisons on left-shifted values into equivalent comparisons on the unshifted variable, adjusting the constant accordingly.
& proposed \\
\hline

8 &
Rewrite conditional (if–then–else) expressions (contained in path condition) into separate cases, one per branch guard, and treat each case as a distinct equivalence class partition
& proposed \\
\midrule
\end{tabularx}
\small proposed = Rule proposed and implemented for this methodology. 
\end{table*}

\section{Examples of Inferred Equivalence Classes}
\label{sec:runningExample}

This section presents a set of illustrative examples showing the ECs inferred from legacy embedded software using the proposed methodology. The examples focus on representative function-level patterns commonly observed in industrial automotive firmware, where the absence of precise specifications complicates systematic EC identification.

Rather than providing a full end-to-end walkthrough of the analysis workflow, the examples highlight the resulting output-oriented equivalence partitions derived from different control-flow and computation patterns. The examples progressively increase in complexity: Example~1 illustrates conditional branching, Example~2 focuses on loop-induced path clustering, and Example~3 combines arithmetic and bit-level operations.

\begin{figure}[ht]
    \centering
    \includegraphics[width=0.9\linewidth]{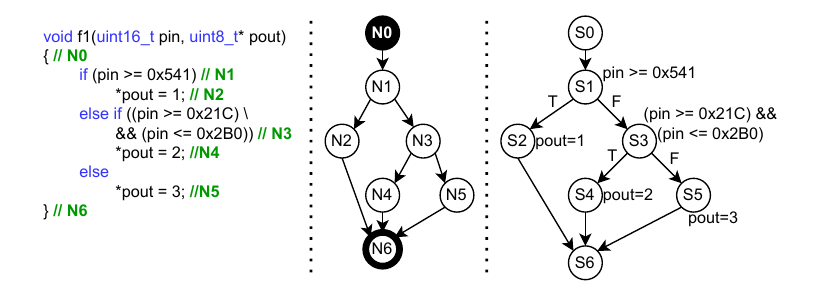}
    \caption{Code, CFG, and symbolic path conditions of \texttt{f1}}
    \label{fig:running_ex1}
\end{figure}

\paragraph{Example~1: Conditional branching.}
We consider a computational function extracted from the executable binary that assigns a value to an output parameter based on a single input variable. The function contains a sequence of conditional branches that partition the input domain into distinct regions, making it a representative example to illustrate output-oriented EC extraction.
Figure~\ref{fig:running_ex1} shows, from left to right, the source code of the function, the reconstructed control-flow graph (CFG), and the symbolic constraints derived from path exploration. While these constraints precisely capture the branching logic, they are expressed in a low-level symbolic form and are not directly suitable for test design.

By grouping paths that induce identical observable effects on the output parameter \texttt{pout}, the methodology derives a compact set of ECs that preserve the original semantics while substantially improving readability. The inferred ECs are post-processed into a human-readable description to support communication and review by test engineers. The resulting ECs and their corresponding input conditions are reported in Table~\ref{tab:tab_EC1}. 

\begin{table}[H]
    \centering
    \caption{ECs derived for \texttt{f1}.}
    \includegraphics[width=1\linewidth]{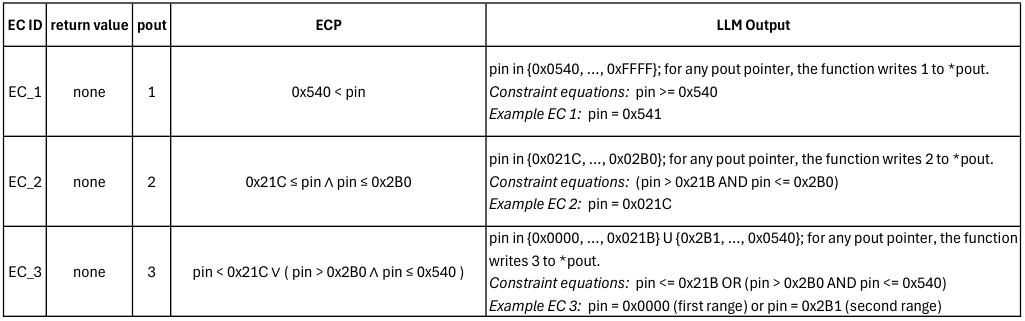}
    \label{tab:tab_EC1}
\end{table}

\begin{figure}[ht]
    \centering
    \includegraphics[width=0.9\linewidth]{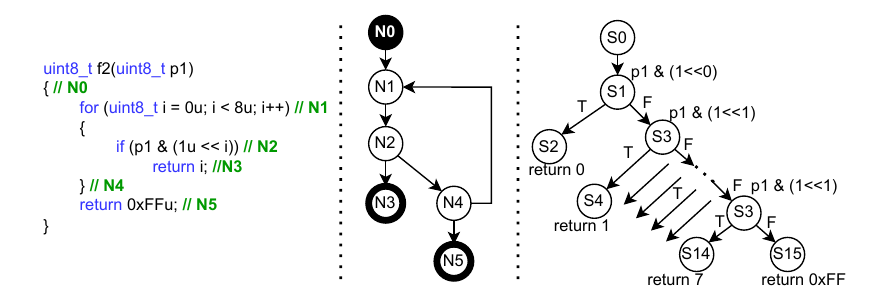}
    \caption{Code, CFG, and symbolic path conditions of \texttt{f2}}
    \label{fig:run_ex2}
\end{figure}

\paragraph{Example~2: Loop-induced path clustering.}
This example considers a function whose control flow is dominated by a loop. The function \texttt{f2} iteratively inspects the bits of its input parameter and returns either the index of the first set bit or a sentinel value when no such bit is found.
Symbolic execution generates multiple execution paths corresponding to different loop iterations. Although these paths differ in the number of iterations, they represent only two distinct observable behaviors: the presence or absence of at least one set bit. Figure~\ref{fig:run_ex2} illustrates the function source code, the reconstructed control-flow graph (CFG), and the symbolic constraints obtained from path exploration.
By grouping loop‑exiting paths with identical observable outcomes, the method collapses all return‑to‑loop paths into a single EC while retaining a distinct class for the case where no bit is set. A final LLM post‑processing step is then applied, as shown in Table~\ref{tab:tab_EC2}.

\begin{table}[H]
    \centering
    \caption{ECs derived for \texttt{f2}.}
    \includegraphics[width=1\linewidth]{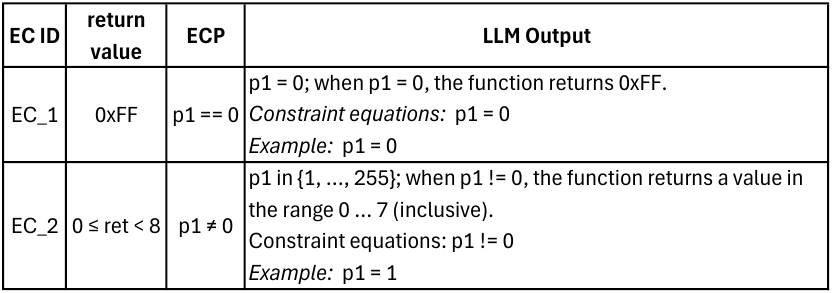}
    \label{tab:tab_EC2}
\end{table}

\begin{figure}[H]
    \centering
    \includegraphics[width=0.9\linewidth]{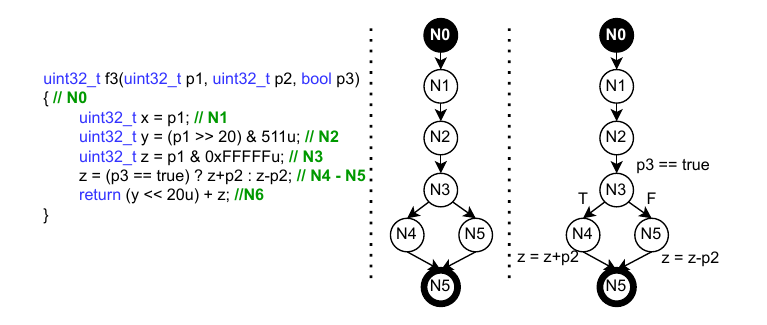}
    \caption{Source code, CFG, and symbolic constraints for \texttt{f3}.}
    \label{fig:run_ex3}
\end{figure}

\paragraph{Example~3: Arithmetic and bit-level constraints.}
This example illustrates equivalence class extraction for a function that combines arithmetic operations with bit-level manipulation of multiple input parameters. The function \texttt{f3} extracts bit fields from its inputs, applies conditional arithmetic based on a boolean flag, and reconstructs the output value.
Different combinations of the boolean flag and arithmetic outcomes give rise to multiple execution paths that, while syntactically distinct, can be grouped according to their observable effects on the returned value. Figure~\ref{fig:run_ex3} shows the function source code, the reconstructed control-flow graph (CFG), and the symbolic constraints derived from path exploration.
Grouping paths with identical interface outputs produces a compact set of ECs encompassing both arithmetic and bit‑level constraints. A post‑processing step then generates human‑readable descriptions, as reported in Table \ref{tab:tab_ec3}.

\begin{table}[H]
    \centering
    \caption{ECs derived for \texttt{f3}.}
    \includegraphics[width=1\linewidth]{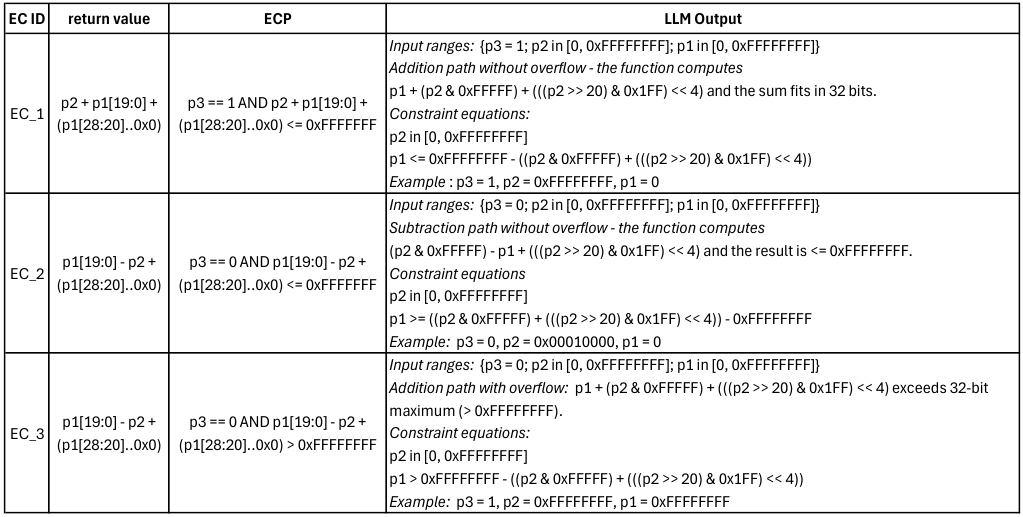}
    \label{tab:tab_ec3}
\end{table}
\section{Experimental Evaluation}
\label{sec:evaluation}

This section presents the industrial validation grounded in the needs identified in Section~\ref{sec:survey}, in particular the lack of systematic support for equivalence class partitioning on legacy firmware and the difficulties of interpreting low-level symbolic constraints. To address these challenges, we conducted a practitioner-centered study involving firmware developers and test engineers, structured to minimize bias and to reflect realistic automotive testing workflows.

Rather than benchmarking the proposed methodology against existing academic techniques, the evaluation focuses on whether the generated equivalence classes are \emph{considered correct, usable, and readable} by practitioners in real-world automotive firmware testing scenarios.

Given the absence of authoritative functional specifications for legacy binaries, correctness is operationalized as agreement with expert-derived expectations of functional behavior. Readability and usability are evaluated separately, with particular attention to the added value of human-readable representations. The validation is organized around the following research questions:

\begin{itemize}[itemsep=0.2em, topsep=0.2em]
\item \textbf{RQ1:} To what extent do the equivalence classes generated by the tool agree with expert-derived expectations of functional behavior? 
\textit{Rationale:} This question assesses the accuracy of the automated extraction by comparing tool-generated partitions against expert expectations in the absence of authoritative specifications.

\item \textbf{RQ2:} To what extent are the generated equivalence classes human-readable and interpretable by firmware engineers? 
\textit{Rationale:} This question evaluates usability by examining whether the inferred classes are readable and suitable for practical test design and review in industrial settings.
\end{itemize}

The research questions are addressed using survey-based \textit{metrics}. Functional correctness, information adequacy, readability, interpretability, and usability are measured through Likert-scale items, while perceived usefulness and missing information are captured through multiple-choice frequency counts. This combination allows both quantitative aggregation and diagnostic insight into practitioners’ perceptions of the generated equivalence classes.
The \textit{objects} of the evaluation are computational functions extracted from industrial automotive firmware binaries. We selected functions implementing pure logic, without direct interaction with peripherals, interrupts, or timing-dependent I/O, and exposing explicit inputs and observable outputs, thus enabling output-oriented equivalence class partitioning. Overall, we considered \textbf{27} functions, with line of code metrics in average 31 LOC. for which the tool generated \textbf{138} equivalence classes, with a median of \textbf{15} and average of \textbf{21} classes per function. These study objects reflect the class of legacy firmware components for which practitioners reported the strongest need for systematic equivalence partitioning support (Section~\ref{sec:survey}).

The \textit{participants} were recruited from the same cohort involved in the preliminary needs assessment described in Section~\ref{sec:survey}. This choice ensured continuity between the industrial needs elicited in the first study and the subsequent validation of the proposed approach, enabling participants to assess the tool outputs against challenges they had previously reported in their daily work.

To address the research questions, we designed a \textit{questionnaire} combining closed-ended items measured on a 5-point Likert scale with multiple-choice questions aimed at collecting categorical and diagnostic feedback. The latter also includes an optional open-ended field for additional comments. Table~\ref{tab:questionnaire} reports the complete questionnaire together with the mapping of items to the corresponding research questions, while the predefined response options for the multiple-choice items are summarized in Table~\ref{tab:multiple-choice-tbl}. The questionnaire is organized into two sections aligned with the two research questions: the first focuses on functional correctness and information adequacy, while the second evaluates readability, interpretability, and perceived utility within typical engineering workflows.

During \textit{execution and analysis}, participants performed a structured evaluation task designed to reflect common testing activities on legacy and undocumented firmware. For each target function, they first analyzed the source code independently and derived equivalence classes according to their usual testing practices, without access to any tool-generated artifacts. They were then presented with the equivalence classes generated by the proposed tool and asked to compare them with their manually derived ones, assessing alignment, completeness, and potential discrepancies. Upon completing this comparison, participants filled in the questionnaire capturing their subjective assessment of the tool-generated artifacts. Survey responses were collected anonymously and analyzed using descriptive statistics. Likert-scale items were aggregated after normalizing reverse-coded questions, while multiple-choice items were analyzed using frequency-based aggregation to synthesize practitioners’ judgments across the considered dimensions.

\begin{table*}[t]
\centering
\small
\renewcommand{\arraystretch}{0.3}
\caption{Questionnaire items and their mapping to the research questions.}
\label{tab:questionnaire}
\begin{tabularx}{\textwidth}{p{0.2cm} X r r}
\toprule
\textbf{ID} & \textbf{Question} & \textbf{Question semantic} &  \textbf{Type} \\
\midrule
\multicolumn{4}{l}{\textbf{Part 1 - RQ1}} \\
\midrule
Q1 &
The produced equivalence classes correctly match the expected functional behavior 
& Functional correctness &  L \\

Q2 &
The produced equivalence classes miss important information, which can make understanding the function behavior difficult 
& Information adequacy &  L \\

Q3 &
The produced equivalence classes contain unnecessary information 
& Information adequacy &  L \\

Q4 &
If there is missing information in the equivalence classes, which kind is it?
& Information adequacy & MC \\

\midrule
\multicolumn{4}{l}{\textbf{Part 2 – RQ2 }} \\
\midrule

Q5 &
The produced equivalence classes are human-readable and interpretable 
& Readability \& Interpretability &  L \\

Q6 &
Which outputs produced by the method do you find useful for understanding the function?
& Practical utility and Usability &  MC \\

Q7 &
Based on your experience, please indicate which of the following statements you believe to be true.
& Practical utility and Usability &  MC \\

Q8 &
Based on my experience in using the method, I find it useful in supporting function behavior comprehension.
& Practical utility and Usability &  L \\

Q9 &
Based on my experience, I am satisfied with the usability of the method.
& Practical utility and Usability &  L \\
\bottomrule
\end{tabularx}
\vspace{0.2em}
\small L = 5-point Likert scale; MC = Multiple choice.
\end{table*}




\begin{table}[]
\centering
\scriptsize
\renewcommand{\arraystretch}{0.7}
\caption{Multiple-choice response options and responses.}
\label{tab:multiple-choice-tbl}
\resizebox{\columnwidth}{!}{
\begin{tabular}{p{6.8cm}c}
\toprule
\textbf{Q4} & \textbf{Votes} \\ 
\midrule
$\Box$ Constraints on symbolic variables. & 1 \\
$\Box$ Boundary conditions. & 6\\
$\Box$ Path feasibility details. & 4\\
$\Box$ Free text for other options. & 3 \\
\midrule
\textbf{Q6} \\ 
\midrule
$\Box$ Equivalence class sets. & 8\\
$\Box$ Simplified constraints. & 7\\
$\Box$ Branch-specific versions. & 4\\
$\Box$ Free text for other options. & 3\\
\midrule
\textbf{Q7} \\ 
\midrule
$\Box$ Compared to manual analysis, the automatically derived equivalence classes enhance understanding. & 10\\
$\Box$ Compared to manual analysis, the automatically derived equivalence classes contain too much information that can interfere with understanding. & 2\\
$\Box$ Manual and automatic approaches provide similar information. & 0\\
$\Box$ Free text for other options. & 0\\
\bottomrule
\end{tabular}}
\end{table}

\subsection{Answer to RQ1}
\label{subsec:results-rq1}

\begin{figure}[ht]
    \centering
    \includegraphics[width=0.9\linewidth]{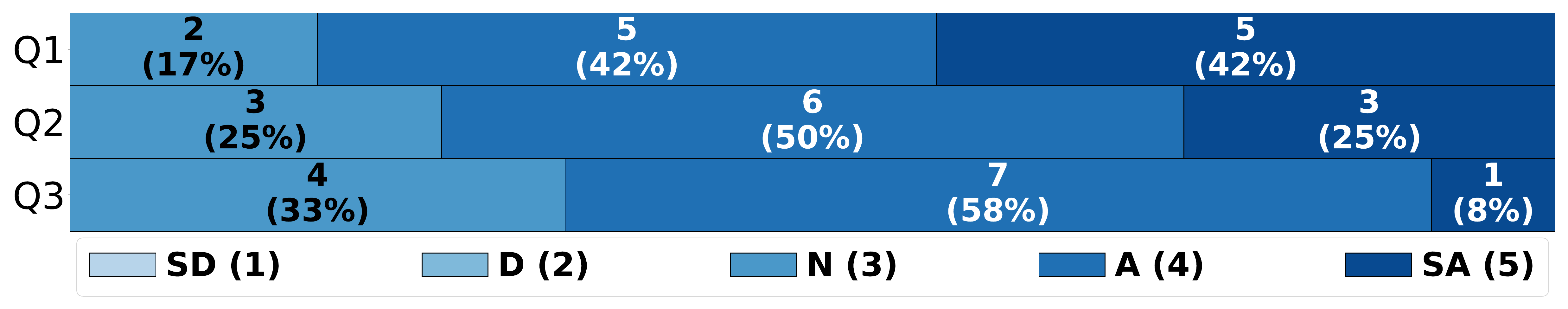}
    \caption{Likert-scale rating distribution for Q1, Q2, Q3}
    \label{fig:rq1-results-Q1-Q2-Q3}
\end{figure}

Figure~\ref{fig:rq1-results-Q1-Q2-Q3} reports the distribution of responses to Q1–Q3, showing an overall positive assessment of the generated equivalence classes in terms of functional correctness and information adequacy. For Q1, which evaluates alignment with expected functional behavior, all responses were positive or neutral, with a clear prevalence of \textbf{Agree} and \textbf{Strongly Agree}, indicating strong consistency between tool-generated results and expert expectations. A similar pattern is observed for Q2, where most respondents expressed agreement, with a limited number of neutral responses, suggesting minor concerns about missing information. Responses to Q3 remain broadly positive, although with slightly higher neutrality, indicating that some participants perceived room for reducing unnecessary detail while still considering the information accurate and appropriate.
Table~\ref{tab:multiple-choice-tbl} summarizes responses to Q4. \textit{Boundary conditions} were the most frequently reported missing aspect, followed by \textit{path feasibility details}, while missing constraints on \textit{symbolic variables} were rarely indicated. Notably, several participants reported \textit{no missing information}. Overall, the identified gaps mainly concern edge-case characterization and path feasibility, rather than the core functional behavior captured by the inferred equivalence classes.


\begin{tcolorbox}[colback=gray!10,boxrule=0.5pt,title=RQ1 Answer,boxsep=1pt,left=1pt,right=1pt,top=1pt,bottom=1pt]
The inferred equivalence classes largely align with expert expectations of functional behavior. Participants reported high perceived correctness, with only minor gaps related to boundary conditions and path feasibility. Overall, the results indicate that the approach produces accurate and sufficiently informative equivalence classes for understanding function behavior.
\end{tcolorbox}

\subsection{Answer to RQ2}
\label{subsec:results-rq2}

\begin{figure}[H]
    \centering
    \includegraphics[width=0.9\linewidth]{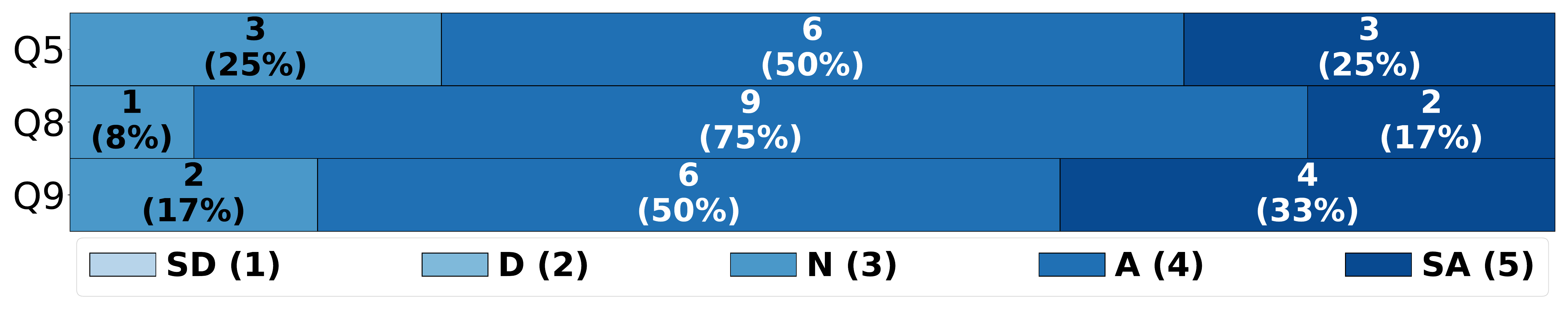}
    \caption{Likert-scale rating distribution for Q5, Q8, Q9}
    \label{fig:rq2-results-Q5-Q8-Q9}
\end{figure}

Figure~\ref{fig:rq2-results-Q5-Q8-Q9} reports responses to Q5, Q8, and Q9, showing a generally positive perception of the generated equivalence classes in terms of readability, interpretability, and usability. Most respondents expressed agreement that the equivalence classes are human-readable and interpretable (Q5), with a limited number of neutral responses indicating minor room for improvement. An even stronger consensus emerges for Q8 and Q9, where the large majority of participants agreed that the approach supports understanding of function behavior and is usable within typical engineering workflows.
Table~\ref{tab:multiple-choice-tbl} summarizes responses to the multiple-choice questions. For Q6, \textit{equivalence class sets} and \textit{simplified constraints} were most frequently selected, indicating a clear preference for compact, high-level representations, while \textit{branch-specific versions} were selected less often. Responses to Q7 further confirm this trend: most participants reported that the generated equivalence classes \textit{enhance understanding} compared to manual analysis, with only a small minority indicating that they contain excessive information.

\begin{tcolorbox}[colback=gray!10,boxrule=0.5pt,title=RQ2 Answer,boxsep=1pt,left=1pt,right=1pt,top=1pt,bottom=1pt]
The inferred equivalence classes are perceived as readable, interpretable, and usable by firmware engineers. Participants reported that the approach supports understanding of function behavior, with clear preference for compact outputs (e.g., equivalence class sets, simplified constraints). Overall, results indicate improved comprehension without introducing excessive information.
\end{tcolorbox}

\subsection{Threats to Validity} 
This study proposes an automated methodology for inferring output-oriented equivalence classes from legacy embedded binaries and evaluates its perceived correctness and usability through an industrial practitioner survey. The main threats to validity and corresponding mitigations are discussed below.

\noindent \textbf{\textit{Construct validity.}}
The correctness of inferred equivalence classes cannot be assessed against complete formal specifications, as the analyzed firmware consists of legacy assets with partially missing or outdated documentation. Practitioners’ judgments are therefore used as a proxy for functional correctness, introducing an inherent degree of subjectivity. To mitigate anchoring effects, the survey adopted a two-phase structure in which participants first analyzed function behavior independently and only then inspected the generated artifacts. Usability and interpretability are assessed through Likert-scale responses and thus reflect perceived utility rather than objective improvements in testing effectiveness.

\noindent \textbf{\textit{Internal validity.}}
The methodology relies on the availability and quality of binary metadata (e.g., DWARF information, map files) to recover function interfaces and control-flow structure. Inaccurate or incomplete metadata may affect constraint reconstruction and, consequently, equivalence class inference. In addition, symbolic execution is sensitive to modeling assumptions and exploration bounds. The implementation adopts constraints typical of safety-critical embedded software (e.g., no recursion, no dynamic memory allocation, bounded loops); deviations from these assumptions or path explosion may result in missed or imprecise behaviors. Since equivalence classes are defined with respect to observable outputs, unmodeled side effects (e.g., hardware interactions, timing-dependent behavior) are not captured.
An optional LLM-based post-processing step is used exclusively to transform already-derived constraints into human-readable form. While it does not affect the symbolic analysis, it may influence perceived readability and is therefore treated strictly as a presentation layer.

\noindent \textbf{\textit{External validity.}}
External validity is limited by both the study objects and the participant sample. Although the relatively small set of analyzed functions may affect generalizability, the methodology has been applied in production environments to a significantly larger corpus of industrial functions. The selected subset was chosen to ensure a manageable workload for practitioners during manual review. The consistent behavior observed across the wider (undisclosed) codebase provides additional confidence in the scalability and robustness of the approach.
Furthermore, the evaluation focuses on computational functions extracted from automotive firmware developed under ISO~26262 constraints and involves a convenience sample from a single organization. As a result, the findings may not generalize to other domains, binary characteristics, or development practices. However, this scope reflects the intended industrial context and aligns with the standard constraints of safety-critical embedded software development.

\section{Lessons Learned from Industrial Adoption}
\label{sec:lessons}

This section discusses lessons learned from the industrial application of the proposed approach, drawing on the experimental evaluation and practitioner feedback. No new empirical evidence is introduced. Given the limited sample size (n=12) and single-organization context, the findings should be interpreted with caution in terms of external validity and generalizability.

\textbf{LL1: Readability is a prerequisite for usability.}
Although the quantitative results indicate a generally positive perception of readability and interpretability (RQ2), practitioners emphasized that relatively minor presentation choices can significantly affect practical usability. Qualitative feedback highlighted formatting aspects and naming practices, noting that \textit{“better indentation would increase readability”} and that \textit{“where possible, numeric literals should be replaced with C macros or named constants.”} These observations suggest that, in industrial settings, equivalence classes are treated as engineering artifacts to be inspected, reviewed, and potentially maintained, rather than as purely analytical outputs. A key lesson is therefore the importance of sensible default formatting (e.g., consistent indentation, whitespace, and visual grouping) and the systematic use of named constants instead of \emph{magic numbers}, supported by lightweight annotations to convey semantics. 

\textbf{LL2: Human-readable and machine-readable outputs serve complementary roles.}
While readability was repeatedly identified as essential for manual inspection and review, the results also indicate that practitioners value representations that strike a balance between human interpretability and the ability to be systematically processed, reused, and integrated into existing industrial workflows. In practice, equivalence classes may thus serve a dual role: supporting human understanding of functional behavior and acting as structured artifacts that enable traceability, automation, and downstream analyses within the development and testing toolchain, particularly in safety-critical contexts where reviewability and traceability are required by standards.
This dual role is further reinforced by unsolicited practitioner comments collected outside the survey, including the preference to \textit{“prefer a structured JSON output, even if it is harder to read.”} Rather than contradicting the emphasis on readability, this feedback highlights the need to balance human interpretability with structural rigor. A practical implication is to support both consumption modes by providing dual outputs: a machine-readable representation suitable for automation and validation, and a concise human-readable summary tailored to review and comprehension tasks.
Overall, these lessons indicate that relatively modest refinements in presentation and output structuring can substantially improve the perceived usability and integration potential of automatically inferred equivalence classes in industrial contexts. At the same time, the findings reinforce that adoption in safety-critical workflows depends not only on analytical correctness but also on predictability, readability, and compatibility with existing toolchains. While preliminary, these insights provide concrete guidance for refining the approach and motivate future studies involving broader populations, multiple projects, and longitudinal evaluations.

\section{Related Studies}
\label{sec:related}

This section briefly reviews existing work on equivalence class partitioning and related specification inference techniques, highlighting their limitations in the context of legacy embedded software.

\textit{Equivalence Class Partitioning} (ECP) is a classical black-box test design technique in which the input domain is divided into subsets expected to exhibit uniform behavior. Traditional \textit{manual ECP} relies on functional specifications, domain knowledge, and heuristics, often combined with Boundary Value Analysis. While effective in well-specified systems, this approach requires substantial human effort and assumes explicit knowledge of input constraints, which is rarely available for large-scale or legacy embedded software. To reduce manual effort, \textit{model-based approaches} automate equivalence class extraction from formal behavioral models such as state machines or transition systems. For example, Hübner et al.~\cite{Hubner2015EquivalenceClass} propose a strategy that is complete with respect to a given fault model and supports infinite input domains. However, these techniques critically depend on the availability and correctness of manually constructed models, which are often unavailable or costly to maintain in industrial embedded contexts.
A large body of work on \textit{specification mining} aims to infer behavioral properties from source code or execution traces to support program comprehension, test generation, and verification. Usage pattern mining techniques capture frequent API call co-occurrences or ordering constraints~\cite{ANAND20131978}, while static and symbolic analyses infer invariants and path conditions through control-flow, data-flow, or symbolic execution. Although symbolic execution enables precise constraint extraction and has been used to recover high-level semantics from binaries~\cite{keliris2018icsref,kim2022reverse,sun2019tell}, existing tools such as DisPatch~\cite{kim2022reverse}, REMaQE~\cite{Udeshi2024}, and PERFUME~\cite{weideman2021perfume} primarily target program understanding or reverse engineering, rather than test-oriented equivalence class extraction.
Declarative approaches based on \textit{Constraint Logic Programming (CLP)} integrate symbolic execution and constraint solving into the execution semantics, enabling systematic path exploration and test case generation~\cite{Albert2014,Gomez2010,Albert2013}. While expressive and compositional, these approaches typically do not treat equivalence classes as first-class testing artifacts and are rarely applied directly to industrial embedded binaries.
Overall, existing approaches exhibit complementary strengths but also limitations that hinder their adoption for equivalence-class-based testing of legacy embedded software. Manual and model-based techniques rely on specifications or models that are often missing in practice, while specification mining and symbolic execution approaches usually treat equivalence classes as a by-product rather than as a primary testing abstraction. The approach proposed in this paper addresses these limitations by directly inferring output-oriented equivalence classes from compiled embedded binaries. By leveraging symbolic execution and constraint merging at the binary level, the method derives compact, test-relevant equivalence partitions without requiring explicit specifications or manually constructed models, aligning symbolic analysis techniques with the practical needs of ISO~26262-compliant testing in industrial settings.

\section{Conclusions and Future Work}
\label{sec:conclusions}

This paper presented a binary-level methodology for inferring equivalence classes from legacy embedded firmware in the absence of reliable functional specifications. The approach targets a common but underexplored industrial scenario in safety domains, where test design techniques are recommended by standards such as ISO~26262. By operating directly on compiled binaries and combining control-flow reconstruction with guided symbolic execution, the method aligns the extracted equivalence classes with the actual deployed firmware configuration.
A central contribution of this work is the treatment of equivalence classes as first-class testing artifacts derived from observable functional behavior. By grouping symbolic execution paths according to equivalent outputs, the approach yields equivalence partitions that support systematic test design rather than merely exposing low-level path conditions. Scalability and industrial feasibility are supported through dependency-aware analysis, bounded symbolic execution, and reusable function summaries, while optional post-processing improves readability without altering the analysis. The industrial evaluation provides preliminary evidence that the inferred equivalence classes align with expert expectations and are considered readable and useful for understanding function behavior and supporting test design in legacy systems. Although the study does not aim at statistical generalization, the results suggest that binary-level equivalence class extraction can effectively complement manual testing in safety-critical industrial contexts.

Future work will focus on extending the approach to broader execution environments and language features, as well as on integrating the extracted equivalence classes into downstream test generation and coverage assessment workflows. Additional empirical studies on larger and more diverse industrial codebases are planned to further assess scalability and practical impact.



\clearpage
\balance
\bibliographystyle{ACM-Reference-Format}
\bibliography{biblo}

\end{document}